\documentclass[english,aps,prb,twocolumn,superscriptaddress,longbibliography]{revtex4-2}
\usepackage[T1]{fontenc}
\usepackage[latin9]{inputenc}
\usepackage{color}
\usepackage{babel}
\usepackage{amsmath}
\usepackage{amssymb}
\usepackage{graphicx}
\usepackage{pgfplots}
\usepgfplotslibrary{groupplots}

\usepackage[unicode=true,pdfusetitle,
bookmarks=true,bookmarksnumbered=false,bookmarksopen=false,
 breaklinks=false,pdfborder={0 0 1},backref=false,colorlinks=false]
 {hyperref}

\usepackage{xcolor}

\definecolor{tri_color}{HTML}{FFB14E}
\definecolor{honey_color}{HTML}{9D02D7}
\definecolor{square_color}{HTML}{0000FF}
\definecolor{corner_color}{HTML}{ea5f94}

\makeatletter
\newcommand{\Tr}{\mathrm{Tr}}
\usepackage{braket}
\usepackage[caption=false]{subfig}
\usepackage{adjustbox}
\newcommand{\myappendlink}{%
  \hyperref[sec:appendix_large_theta]{Appendix}%
}

\makeatother

\begin{document}
\global\long\def\k#1{\Ket{#1}}%
\global\long\def\b#1{\Bra{#1}}%
\global\long\def\bk#1{\Braket{#1}}%
\global\long\def\Tr{\mathrm{Tr}}%
\global\long\def\Var{\text{Var}}%

\title{Corner contribution to the entanglement in two-dimensional systems: A tensor network perspective}
\author{Noa Feldman}
\author{Moshe Goldstein}
\affiliation{Raymond and Beverly Sackler School of Physics and Astronomy, Tel-Aviv University, Tel Aviv 6997801, Israel}
\begin{abstract}
In continuous quantum field theories, the entanglement entropy of a subsystem with sharp corners on its boundary exhibits a universal corner-dependent contribution. We study this contribution through the lens of lattice discretization, and demonstrate that this corner dependence emerges naturally from the geometric structure of infinite projected entangled pair states (iPEPS) on discrete lattices. Using a rigorous counting argument, we show that the bond dimension of an iPEPS representation, which serves as a bound on the entanglement, exhibits a corner-dependent term that matches the predicted term in gapped continuous systems.
Crucially, we find that this correspondence only emerges when averaging over all possible lattice orientations and origin positions, revealing a fundamental requirement for properly discretizing continuous systems. Our results provide a geometric understanding of entanglement corner laws and establish a direct connection between continuum field theory predictions and the structure of discrete tensor network representations. We extend our analysis to gauge-invariant systems, where lattice corners crossed by the bipartition boundary contribute an additional corner-dependent term. These findings offer new insights into the relationship between entanglement in continuous and discrete quantum systems.
\end{abstract}

\maketitle

\section{introduction}
Tensor networks (TNs), specifically projected entangled pair states (PEPSs)~\citep{verstraeteRenormalizationalgorithmsquantummanybody2004}, are extremely useful in representing a vast range of physical states on discrete lattices. Their translationally-invariant version, infinite PEPSs (iPEPSs)~\citep{VerstraeteIPEPS2008,JordanIPEPS2008,verstraeteRenormalizationalgorithmsquantummanybody2004}, has proven useful in the study of ground states~\citep{PicotNematicSupernematic2015,CorbozImprovedEnergyExtrapolation2016,KshetrimayumSpin12Kagome2016,PicotSpinSKagome2016,LiaoGapplessSpinLiquid2017,BoosCompetitionIntermediatePlaquette2019,KSHETRIMAYUM2020,AstrakhantsevPinwheel2021}, finite temperature behavior~\citep{thermal_LiLinearizedTRG2011,thermal_CzarnikPEPSFiniteTemperature2012,thermal_DongBilayer2017,thermal_KshetrimayumTNAnnealing2019,thermal_CzarnikKitaevHeisenberg2019,thermal_MondalTwoBody2020,thermal_CzarnikTNStudym12Magnetization2021,thermal_SchmollFiniteTemperature2024}, and nonequlibrium~\citep{neq_KshetrimayumSteadyStates2017,neq_CzarnikTimeEvolutionIPEPS2019,neq_HubigTimeDependentDisordered2019,BoosCompetitionIntermediatePlaquette2019,neq_KshetrimayumMBL2020,neq_KshetrimayumTimeCrystals2021,neq_DziarmagaMBLTimeCrystals2022,neq_DziarmagaGradientTensorUpdate2022,FirankoAreaLawSteadyStates2024} of  various open problems, and they serve as a state-of-the-art numerical tool for the study of systems that suffer from the sign problem~\citep{LohSignProblem1990}. 

The explicit relation between the TN representation of a state and its entanglement scaling make TNs a useful analytical tool for the study of entanglement in many-body states (see, e.g., Refs. ~\citep{FeldmanSuperselection2024,KnauteEntanglementConfinement2024} on the entanglement of gauge-invariant PEPSs). 
As explained in detail in Sec. \ref{sec:entanglement_in_ipeps} below, the property referred to as \textit{bond dimension} of a TN-represented state serves as an upper bound to the Schmidt rank (zeroth R\'enyi entropy) of the state, and in turn, also to the von Neumann entropy. Thus, assessing the behavior of the bond dimension allows us to probe the entanglement of TN-representable states. Focusing on 2D-iPEPS-representable states, we study here two-dimensional systems with area-law entanglement (as a leading term in the entanglement scaling).

We summarize our results briefly: We study the entanglement of a subsystem $A$  of a two-dimensional system, with a sharp corner $\theta$ in its boundary, as depicted in Fig. \ref{fig:bipartition}a. Refs.
~\citep{corner_Selem2013,corner_cft_Bueno_2015,corner_cylinder_Rodriguez_2010,corner_Crepel2021,corner_general_Estienne_2022,corner_Bueno2023,corner_DEmidio2024} 
show that in continuous field systems, the entanglement of $A$  and its environment is of the form 
\begin{equation*}
    S(A) = b_\text{vol}|A| + b_\text{area} |\partial A| - b_A\left(\theta,|A|,|\partial A|\right),
\end{equation*}
where $S(A)$  measures the entanglement of $A$  with its environment (defined in Eq. (\ref{eq:vN})), $|A|, |\partial A|$  are $A$'s bulk (volume) and boundary (area) sizes, respectively, and $b_\text{vol},b_\text{area}$  are constants. Note that, typically, in the ground state of gapped or critical (gapless) systems, the entanglement scales as an area-law, i.e., $b_\text{vol} = 0$, and the leading term of the entanglement is proportional to $|\partial A|$~\citep{FradkinMoore2006,Eisert_area_law}. $b_A(\theta)$  is a subleading term representing the dependence of the entanglement on $\theta$, and may generally depend on  $|A|, |\partial A|$. Refs.~\citep{corner_cft_Bueno_2015,corner_general_Estienne_2022} calculate it in several cases: In several gapped condensed matter systems such as fractional quantum Hall (FQH) systems, $b_A(\theta,|A|,|\partial A|)=b_A(\theta)$. The corner-dependent behavior of continuous systems is further studied in Refs.~\citep{corner_charge_flucts_Berthiere2023,corner_Crepel2021,wu2024cornerchargefluctuationsmanybody}.

In our work, we use the iPEPS structure to show rigorously that in an iPEPS-represented state on a square lattice, a $\theta$-dependent term in the bond dimension term emerges, which does not depend on $|A|, |\partial A|$. We show that, upon averaging over the lattice orientation and origin position, the behavior of the emergent term approximates the dependence of $b_A(\theta)$  in gapped continuous systems, as obtained in Refs.~\citep{corner_cft_Bueno_2015,corner_general_Estienne_2022}.  This is done by a simple counting argument.  The total bond dimension of a subsystem of the iPEPS bounds the Schmidt rank of the reduced density matrix, and thus, in turn, bounds entanglement measures from above. Moreover, often the scaling of the total bond dimension with the system parameters is often similar to that of entanglement measures. We thus connect the analytical field theory arguments with the geometric properties of their discretized versions.

The rest of this paper is organized as follows: In Sec. \ref{sec:entanglement_in_ipeps}  we present background on the definition and properties of PEPSs, the entanglement measures used in this work, and the behavior of entanglement in PEPS-represented states. We then present the previously-predicted corner-dependent term in Sec. \ref{sec:corner-law}, and discuss its emergence on iPEPS-represented states. We then show, based on TN properties, that the corner-dependent term as predicted for gapped systems emerges in iPEPSs, with the full technical details relegated to the \myappendlink. 
\textcolor{black}{We demonstrate our argument on general systems, as well as gauge-invariant systems in which the entanglement depends on the lattice corners crossed by the boundary~\citep{FeldmanSuperselection2024}.}
\textcolor{black}{ We conclude in Sec. \ref{sec:conclusion}.}

\section{Entanglement in iPEPS and iMPS\label{sec:entanglement_in_ipeps}}

\begin{figure}
    \centering
    \includegraphics[width=\linewidth]{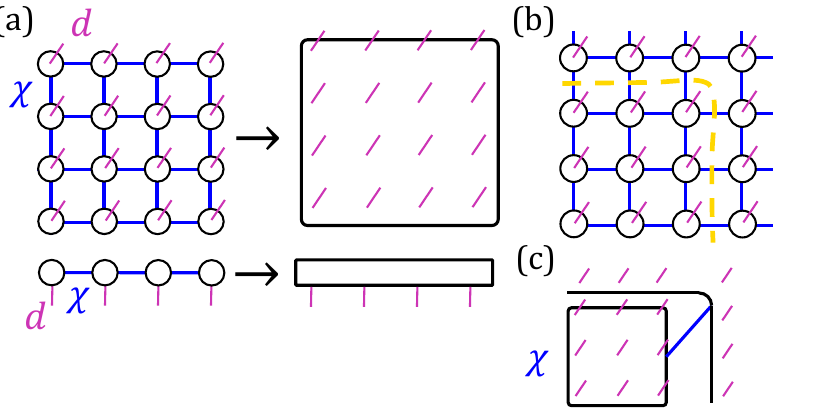}
    \caption{(a) A TN is composed of tensors, each corresponding to a lattice site. The leg colored in pink of each tensor is referred to as the \textit{physical leg}, and corresponds to an index varying over a basis of the Hilbert space of a single lattice site, while legs colored in blue are referred to as the \textit{virtual legs}, corresponding to summed-over indices corresponding to a virtual Hilbert space with dimension $\chi$, referred to as the \textit{bond dimension}. A two- (or higher) dimensional TN is referred to as a PEPS, and a one-dimensional TN is referred to as an MPS. (b) The number of nonzero RDM eigenvalues is bounded from above by the total bond dimension across the boundary between the two subsystems, here depicted in yellow. (c) The bond indices on the boundary may be merged into a single index, whose dimension is the product of all boundary bond dimensions. This dimension bounds the rank of the Schmidt decomposition, Eq. (\ref{eq:schmidt}).}
    \label{fig:tn_intro}
\end{figure}

The PEPS ~\citep{verstraeteRenormalizationalgorithmsquantummanybody2004} representation of a two-dimensional square lattice state with $n$ sites and local Hilbert space size $d$ is composed of a two-dimensional array of rank-5 tensors, each tensor corresponding to a lattice site. One index of each tensor, referred to as \textit{the physical leg}, corresponds to the local Hilbert space of the lattice site or unit cell represented by the tensor, while the other four indices, referred to as \textit{the virtual legs}, correspond to a virtual Hilbert space of dimension $\chi$  and represent the correlation between the lattice sites. $\chi$ is referred to as \textit{the bond dimension} of the tensor. When contracting the virtual legs of neighboring tensors as in Fig. \ref{fig:tn_intro}a, a $d^{\otimes n}$ tensor is obtained, which can be reshaped into a $d^n$ state vector, the standard representation of a quantum state.
The one-dimensional analogue of this representation is called an MPS~\citep{SCHOLLWOCK201196}, as illustrated in Fig. \ref{fig:tn_intro}a.

Special cases of PEPSs and MPSs are the iPEPS~\citep{VerstraeteIPEPS2008,JordanIPEPS2008,verstraeteRenormalizationalgorithmsquantummanybody2004} and iMPS~\citep{Xiang_imps_2023}: The state is represented by a single PEPS or MPS tensor, repeated  over the lattice and resulting in a translationally invariant state over an infinite system. An important property of iPEPSs and iMPSs used in this work is that the bond dimension $\chi$ is constant across the lattice.

TN representations are inherently related to the entanglement of the state, as we show below. We recap the definition of standard entanglement measures for a system in a pure state $\k{\psi}$. They quantify the entanglement between a subsystem $A$  and its complement $B=\overline{A}$. The reduced density matrix (RDM) of $A$ is defined by 
\begin{equation}
    \label{eq:RDM}
    \rho_A = \Tr_B \k{\psi}\b{\psi},
\end{equation}
where $\Tr_B[\cdot]$ stands for a partial trace over the degrees of freedom of $B$. The primary measure of entanglement between $A$ and $B$  is the von Neumann entropy, 
\begin{equation}
    \label{eq:vN}
    S(A) = -\Tr \left[\rho_A \log \rho_A \right] = -\sum_\lambda \lambda\log\lambda,
\end{equation}
where $\lambda$ are the RDM eigenvalues. Additional, numerically accessible measures, are the R\'enyi entropies:
\begin{equation}
    \label{eq:2nd_renyi}
    S^{(n)}(A) = -\frac{1}{1-n}\log\left(\Tr\left[\rho_A^n\right]\right) = -\frac{1}{1-n}\log\left(\sum_\lambda \lambda^n\right).
\end{equation}
From Eqs. (\ref{eq:vN}), (\ref{eq:2nd_renyi})  it is apparent that the entanglement is bounded by the log of the number of nonzero RDM eigenvalues $\lambda$ (its Schmidt rank). 

We now bound the entanglement between a subsystem $A$ and its complement in a PEPS-represented state, relying on the Schmidt decomposition~\citep{SCHOLLWOCK201196}: Any state vector of the total system may be written in the form 
\begin{equation}
    \label{eq:schmidt}
    \k{\psi}_{AB} = \sum_{i=1}^N \psi_i \k{i}_A \k{i}_B,
\end{equation}
where $\{\k{i}_A\}_{i=1}^N, \{\k{i}_B\}_{i=1}^N, $ are orthonormal states of subsystems $A, B$, respectively, not necessarily spanning fully the Hilbert spaces $\mathcal{H}_A,\mathcal{H}_B$. \textcolor{black}{$N$  is referred to as \textit{the Schmidt rank}, and} $\psi_i\ne 0$  are referred to as the \textit{Schmidt values} of $\psi$. In the basis above, the RDM of $A$  is obtained by
\begin{equation}
    \label{eq:schmidt_rdm}
    \rho_A = \sum_{i=1}^N |\psi_i|^2 \k{i}_A\b{i}_A,
\end{equation}
and it is apparent that the Schmidt values squared are the RDM eigenvalues. Therefore, the number of nonzero Schmidt values $N$ is the number of nonzero RDM eigenvalues, bounding the entanglement. 

This formalism decomposes the state into the local Hilbert spaces of $A$ and $B$, similarly to the local decomposition of a PEPS state. In a PEPS, number of virtual legs connecting the tensors corresponding to subsystems $A,B$  equals the boundary (area) between the systems, as illustrated in Fig. \ref{fig:tn_intro}b.  Contracting the virtual legs correlating sites inside the same subsystem, and merging all of the connecting virtual legs, as depicted in Fig. \ref{fig:tn_intro}c, results in a representation of the state in the decomposed Hilbert space. This demonstrates the relation of the virtual bond dimension and the entanglement between $A,B$:  The Schmidt rank $N$,  and in turn, the entanglement,  are bounded by the bond dimension of the merged virtual leg.  In iPEPS- and iMPS-represented states this results in an area-law scaling of the entanglement, $N\le\chi^{|\partial A|}$ (which is constant in the one-dimensional case). Note that $N$  is only bounded by the total bond dimension $\chi^{|\partial A|}$  and is not equal to it, since the iPEPS representation does not necessarily correspond to the Schmidt form (the respective states of each subsystem are not orthonormal in general).  This property holds also in the case where the nodes break spatial symmetries. In our analysis below, we show that in extracting the corner-dependent term, averaging over the directionality of the lattice is necessary.

\section{Field theory corner-law entanglement from iPEPS \label{sec:corner-law}}

\begin{figure}
    \centering
    \includegraphics[width=1\linewidth]{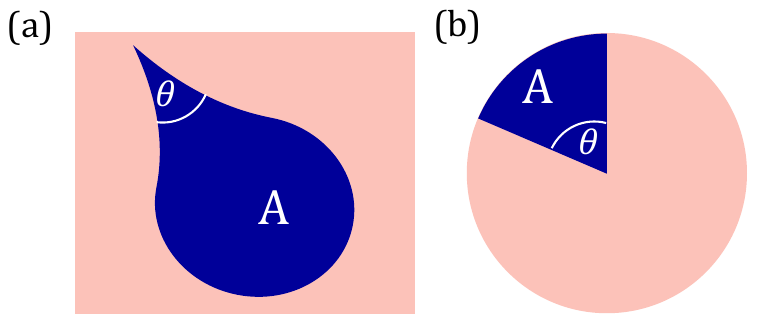}
    \caption{(a) The studied bipartition between a subsystem $A$  and its complement, where the bipartition boundary contains a sharp corner $\theta$. (b) Following Ref.~\citep{corner_general_Estienne_2022}, the results are obtained for a circular system of radius $r$ with subsystem $A$ being a circular section with central angle $\theta$. \textcolor{black}{The results are assumed to hold for an infinite system, that is, $r\rightarrow\infty$.} }
    \label{fig:bipartition}
\end{figure}
In this section,  we study the $\theta$-dependent term in the Schmidt rank of $\rho(A)$, which also relates to the entanglement behavior. We use a counting argument to show that such a term behaves similarly to the behavior of $b(\theta)$  in continuous systems, and approaches the same limits for both $\theta\rightarrow0$ and $\theta\rightarrow\pi$. 
The analysis is presented for a square lattice, for simplicity, but can be straightforwardly shown to generalize naturally to other lattice types, as is demonstrated in Fig. \ref{fig:counting_numerics} below.

We start by describing the expected term for continuum fields:  Suppose a two-dimensional system is partitioned into subsystems $A,B$, and that the bipartition's geometry contains a sharp corner of angle $\theta$, as depicted in Fig. \ref{fig:bipartition}a. Refs.~\citep{corner_cft_Bueno_2015,corner_general_Estienne_2022}  show that the charge fluctuations between $A,B$  behaves as
\begin{equation}\label{eq:scaling_terms}
    S(A) = b_\text{vol} |A| + b_\text{area} |\partial A| - b_A\left(\theta, |A|, |\partial A|\right),
\end{equation}
where $|A|,|\partial A|$  correspond to the ``volume'' (area in 2D) and boundary ``area'' (length in 2D) of $A$, respectively, and $b_A$ depends on $\theta$ and possibly other features of $A$ as well. Ref.~\citep{corner_general_Estienne_2022} finds the universal dependence of $b$ on $\theta$:
\begin{equation}\label{eq:theta_dependence}
b_A(\theta) = -(1 + (\pi-\theta)\cot(\theta)) \cdot F(|A|,|\partial A|),
\end{equation}
where $F$ depends on the model properties. 
This dependence is obtained by studying the entanglement in a pure circular system, where the bipartition is composed of two edges emanating the center with an angle $\theta$ between them, as depicted in Fig. \ref{fig:bipartition}b.

$F$ is studied in Ref.~\citep{corner_general_Estienne_2022} for various specific models: In fractional quantum Hall (FQH) and additional gapped states, it is shown that $F$ is independent of  $|A|,|\partial A|$:
\begin{equation}\label{eq:fqh_term}
    b_A(\theta)  =  -(1 + (\pi-\theta)\cot(\theta)) \cdot \text{const.},
\end{equation}
while in critical systems, $F$ is found to depend logarithmically on $|\partial A|$:
\begin{equation}\label{eq:cft_term}
    b_A(\theta) \propto -(1 + (\pi-\theta)\cot(\theta)) \cdot \log\left(|\partial A|\right).
\end{equation}
This behavior is rigorously established in Ref.~\citep{corner_general_Estienne_2022} for charge fluctuations, and is anticipated to generalize to entanglement properties as well. This is supported by the numerical evidence presented in the same reference. Subsequently, Ref.~\citep{corner_charge_flucts_Berthiere2023} provides an analytical treatment of the corner-angle dependence for general state properties (referred to as charge cumulants), including entanglement measures. Their analysis reveals that for small opening angles $\theta$, the scaling $b_A(\theta)\sim1/\theta$ emerges, which is consistent with Eq.~(\ref{eq:theta_dependence}).

We now show that the behavior in FQH systems, Eq. (\ref{eq:fqh_term}) may be reproduced in iPEPSs in the continuum limit, at least approximately. Our argument relies on a simple calculation of the dependence of the total bond dimension on $\theta$, which in turn, bounds the $\theta$-dependent term in the entanglement, as explained in Sec. \ref{sec:entanglement_in_ipeps} above.

\begin{figure}
    \centering
    \includegraphics[width=\linewidth]{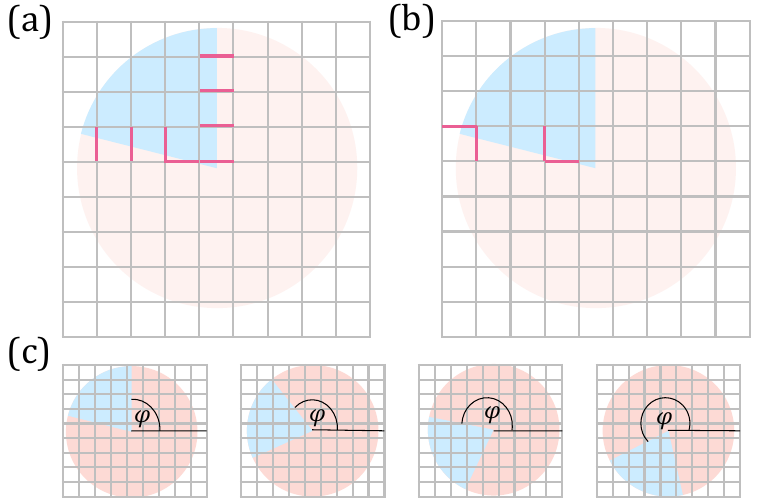}
    \caption{(a) The gray grid indicates the discretized lattice, with the system's degrees of freedom residing on the lattice sites. The virtual legs crossed by the bipartition are accented in pink, their total number is denoted by $n_\text{legs}$. (b) Lattice corners crossed by the bipartition are accented in pink, their total number is denoted by $n_\text{corners}$. (c) In order to conserve the rotational invariance of the system, we average $n_\text{legs},n_\text{corners}$  over different orientations of subsystem $A$. The orientation angle is denoted throughout the paper by $\varphi$.}
    \label{fig:discretization}
\end{figure}

We follow Ref. \citep{corner_general_Estienne_2022} and study the bipartition in Fig. \ref{fig:bipartition}b, where the full system is a circle with radius $r$, and $A$  is a circular section with a central angle $\theta$. One may now bound the entanglement $S(A)$ by bounding the total bond dimension crossed by the bipartition, which, in an iPEPS, equals $\chi^{n_\text{legs}}$, where $n_\text{legs}$ is the number of virtual legs crossed by the boundary of $A$ (See Fig. \ref{fig:discretization}a). We note that the fact that $n_\mathrm{legs}$ has a corner contribution was mentioned, though not calculated, for charge fluctuations of free-fermion systems~\citep{PokManTamCornerCharge2024}. Additionally, we follow Ref.~\citep{FeldmanSuperselection2024}, which shows that in gauge-invariant systems, one observes a lattice-corner-law behavior in the entanglement, that is, a contribution to entanglement measures which scales as the number of lattice corners in the system (to be distinguished from the bipartition corners, which are well-defined in the continuum limit, see Fig. \ref{fig:discretization}b). We therefore also consider separably the number of lattice corners crossed by the bipartition's boundary, $n_\text{corners}$. In order to recover the rotational symmetry of continuous systems, we average over all possible orientations of the subsystem $A$ with respect to the lattice, denoted by $\varphi$, as illustrated in Fig. \ref{fig:discretization}c. The position of the vertex of the angle $\theta$ is averaged over a single unit cell of the lattice. The results are then fitted, based on Eq. (\ref{eq:theta_dependence}), to 
\begin{equation}\label{eq:fit_function}
    f(\theta) = \alpha +\beta (\pi - \theta)\cot\theta.
\end{equation}

\begin{figure}
    \centering
    \includegraphics[width=\linewidth]{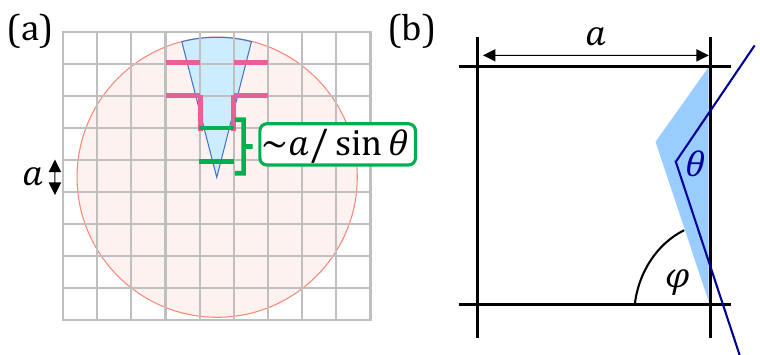}
    \caption{(a) The number of virtual legs $n_\text{legs}$ crossed by the boundary of $A$ when $\theta\rightarrow 0$ . Legs crossed by the boundary are colored in pink, while legs ``skipped'' by the boundary are colored in green. The number of legs skipped depends on the orientation and vertex position.
 (b) When $\theta > 2\tan^{-1}(1/2)$, the only legs that may be ``skipped'' are the ones closest to the apex, as depicted in the illustration. For a given orientation $\varphi$, the apex positions for which the leg will be ``skipped'' are colored in blue. In the \myappendlink \ we integrate over this blue region and over $\varphi$  to obtain $p_\text{skip}(\theta)$. }
    \label{fig:bond_dim_analysis}
\end{figure}

In Fig.~\ref{fig:bond_dim_analysis}  we show how the $\theta$ dependence from Eq. (\ref{eq:theta_dependence}) is recovered for an iPEPS in the limits $\theta \rightarrow 0$ and $\theta \rightarrow \pi$. The dependence on $\theta$ emerges in vicinity of its apex: The number of virtual legs crossed by the boundary far away from the apex (colored in pink in Fig. \ref{fig:bond_dim_analysis}a) is independent of $\theta$ once averaged over different orientations of $A$. 
When $\theta\rightarrow0$, legs close to the apex that could have been crossed by the boundary if $\theta$  was larger, may be ``skipped'' due to $\theta$'s sharpness (green legs in Fig. \ref{fig:bond_dim_analysis}a). The number of green legs does not depend on  $|\partial A|\sim r$. It scales roughly as $a/\sin\theta$, where $a$ 
 is the lattice spacing. In the limit $\theta\rightarrow0$, this behavior matches the $\theta$-dependent term in $f(\theta)$. While this limit may not be taken fully, since $\tan\theta$  is required to be larger than $a/r$,  the $\theta$-dependence of $f$  is already observed as $\theta$  approaches its minimal value. Similar analysis may be applied to $n_\text{corners}$. 

 In the limit $\theta\rightarrow\pi$, or in fact, already for $\theta>2\tan^{-1}(1/2)$, the virtual legs ``skipped'' by the boundary of $A$ may only stem from the legs closest to the apex, when the apex is close enough to one of the edges of the unit cell, as may be seen in Fig. \ref{fig:bond_dim_analysis}b. We denote the probability that the unit cell bond may be ``skipped'' by $p_\text{skip}(\theta)$. In the \myappendlink \ we perform the full analysis of $p_\text{skip}(\theta)$, and find that it is proportional to $\frac{a^2}{16}(1-(\pi-\theta)\cot \theta)$, exactly as in $f(\theta)$. A similar analysis is also applied there to $n_\text{corners}$.
 
In both cases, the result does not depend on the ``area'' $|\partial A|$, i.e., $\beta=\text{const}$. This is true both for the number of virtual legs as well as for the number of lattice corners that reside on the boundary. 

The counting argument presented above connects the continuum field theory entanglement corner term with the geometric structure of lattice systems. Although this argument can be extended to other lattice types, it consistently relies on averaging over lattice orientations. This necessity suggests that certain features inherent to continuous systems may be wrongfully represented when a fixed lattice orientation is used. 

Another important point to consider is the case of critical systems, where Eq.~(\ref{eq:cft_term}) shows that $b(\theta) \sim \log|\partial A|$. The $\theta$-dependence of the bond dimension derived above is independent of $|\partial A|$, and determined solely by a fixed region near the corner, with a radius proportional to $\theta$, denoted by $r_\theta$. Therefore, our counting argument does not account for the $\theta$-dependence observed in critical systems. 
This is not a contradiction, since our argument provides an upper bound on the Schmidt rank, and the corner term is negative. Moreover, when considering other entanglement measures, a corner term with logarithmic scaling in $|\partial A|$ may arise from the values of the singular values themselves, rather than merely from the count of nonzero singular values, as we do here. A possible option for obtaining a geometric log dependence on the area could be to use a different TN structure, e.g., MERA~\citep{MERAVidal2007,MERAVidal2008,MERACincio2008,MERAEvenbly2009,MERAEvenbly22009} Another option is a discretized system that lives on a fractal rather than a lattice. A fractal system will also conserve the scale invariance that is inherent to critical systems. It would be interesting to try and generalize the idea we use here to critical systems discretized on a fractal, and study what other features may be obtained this way.

\subsection{Numerical demonstrations}

In addition to the analysis of $f(\theta)$  for $\theta\rightarrow 0$ and $\theta\rightarrow\pi$, we perform a numerical count of $n_\text{legs}, n_\text{corners}$ as a function of $\theta$ for $\theta\in[0,\pi]$, for varying values of $r$.
We use exact counting of virtual legs and lattice corners crossed by the bipartition boundary for three lattice types: square lattice, where we count both $n_\text{legs}$ and $n_\text{corners}$, and triangular and honeycomb lattices, where we only count $n_\text{legs}$. The relative variation of $\sim 10^{-2}$ in the extracted coefficients $\beta_\text{legs}$ for all lattice types and $\beta_\text{corners}$ arises from the least-squares fitting procedure used to obtain $\beta$ from the discrete counts. The fits to Eq. (\ref{eq:fit_function}) were performed over the full range $\theta \in [0, \pi]$, with typical error remaining below $10^{-5}$  for large $r$, as shown in Fig. \ref{fig:counting_numerics}c. 

In Fig. \ref{fig:counting_numerics} we show, as an example, the dependence of $n_\text{legs}$ on $\theta$  for a particular value of $r$, fitted to Eq. (\ref{eq:fit_function}). We then fit $n_\text{legs}, n_\text{corners}$  to Eq. (\ref{eq:fit_function}) for various values of $r$. We plot the dependence, or rather the lack thereof, of the extracted $\beta_\text{legs},\beta_\text{corners}$ on $r$, demonstrating the independence of $b(\theta)$ on $|\partial A|$. The coefficients $\beta_\text{legs},\beta_\text{corners}$ are constant as a function of the area, up to a relative variation of $\sim10^{-2}$, which may account for finite-size effects or numerical errors.  In order for the dependence in Eq. (\ref{eq:fqh_term}) to be accounted for, $\alpha$ may and is in fact expected to depend on $|\partial A|$. 

We may interpret these results from a different point of view: We may take $|\partial A| = \text{const.}$, and consider increasing $r$ to be the result of decreasing the lattice constant $a$, thus fine graining the system, with the appropriate normalization. $r\rightarrow\infty$  thus corresponds to the continuum limit (rather than an infinite system). As may be seen in Fig. \ref{fig:counting_numerics}, the fit to Eq. (\ref{eq:fit_function}) becomes better as $r$ grows larger, i.e., as the continuum limit is approached. All of the above indicates that $b(\theta)$, the corner-dependent term in the entanglement, is independent of $|\partial A|$, and that Eq. (\ref{eq:fqh_term}) is recovered.

We stress that the bond dimension $\chi$  typically grows with the accuracy as the continuum limit is approached, but as long as it does not depend on the size of subsystem $A$, our argument applies. The corner-dependent term of the total bond dimension (and in turn, the number of nonzero RDM eigenvalues) does not depend on the area, as long as the bond dimension $\chi$  is uniform across the lattice. 

\begin{figure}
\raggedright
(a)
  \begin{minipage}{0.5\textwidth}
    \centering
    % ===== Plot 1: Bond count vs theta at r=60 with fits =====
\begin{tikzpicture}
\begin{axis}[
    xlabel={$\theta$},
    ylabel={Average bond count},
	legend style={at={(1,1.01)}, anchor=south east},
    legend cell align=left,
    legend columns=4,
    grid=major,
    xmin=0, xmax=3.2,
    xtick={0, 0.7854, 1.5708, 2.3562, 3.1416},
    xticklabels={$0$, $\pi/4$, $\pi/2$, $3\pi/4$, $\pi$},
]

% Square data
\addplot[only marks, mark=*, mark size=1.5pt, square_color]
table[x=theta, y=count, col sep=comma]{fit_bonds/square_r60_data.csv};
\addlegendentry{Square}

% Square fit
\addplot[square_color!40!black, dashed, thick, no markers, forget plot]
    table[x=theta, y=fit, col sep=comma]{fit_bonds/square_r60_fit.csv};
% \addlegendentry{Square fit}

% Triangular data
\addplot[only marks, mark=*, mark size=1.5pt, tri_color]
    table[x=theta, y=count, col sep=comma]{fit_bonds/triangular_r60_data.csv};
 \addlegendentry{Triangular}

% Triangular fit
\addplot[tri_color!60!black, dashed, thick, no markers, forget plot]
    table[x=theta, y=fit, col sep=comma]{fit_bonds/triangular_r60_fit.csv};
% \addlegendentry{Triangular fit}

% Honeycomb data
\addplot[only marks, mark=*, mark size=1.5pt, honey_color]
    table[x=theta, y=count, col sep=comma]{fit_bonds/honeycomb_r60_data.csv};
 \addlegendentry{Honeycomb}

% Honeycomb fit
\addplot[honey_color!60!black, dashed, thick, no markers, forget plot]
    table[x=theta, y=fit, col sep=comma]{fit_bonds/honeycomb_r60_fit.csv};
% \addlegendentry{Honeycomb fit}

% Square corners data
\addplot[only marks, mark=*, mark size=1.5pt, corner_color]
    table[x=theta, y=count, col sep=comma]{fit_bonds/square_corners_r60_data.csv};
 \addlegendentry{Square corners}

% Square corners fit
\addplot[corner_color!60!black, dashed, thick, no markers, forget plot]
    table[x=theta, y=fit, col sep=comma]{fit_bonds/square_corners_r60_fit.csv};
% \addlegendentry{Square corners fit}

\end{axis}
\end{tikzpicture}
    \hfill 
\end{minipage}
(b)
  \begin{minipage}{0.5\textwidth}
    \centering
    \begin{tikzpicture}
\begin{axis}[
    height=4cm, width=\linewidth,
    xlabel={$r$},
    ylabel={$\beta/\overline{\beta}$},
    legend pos=south east,
    legend cell align=left,
    grid=major,
    scaled y ticks=false,
    yticklabel style={
        /pgf/number format/fixed,
        /pgf/number format/precision=3, % Adjust as needed for your data
        /pgf/number format/fixed zerofill % Ensures consistent decimal length
    },
]

\addplot[square_color, mark=*, thick]
    table[x=r, y=square, col sep=comma]{fit_bonds/beta_vs_r.csv};
% \addlegendentry{Square}

\addplot[tri_color, mark=*, thick]
    table[x=r, y=triangular, col sep=comma]{fit_bonds/beta_vs_r.csv};
% \addlegendentry{Triangular}

\addplot[honey_color, mark=*, thick]
    table[x=r, y=honeycomb, col sep=comma]{fit_bonds/beta_vs_r.csv};
% \addlegendentry{Honeycomb}

\addplot[corner_color, mark=*, thick]
    table[x=r, y=square_corners, col sep=comma]{fit_bonds/beta_vs_r.csv};
% \addlegendentry{Square corners}

\end{axis}
\end{tikzpicture}
    \hfill 
\end{minipage}
(c)
  \begin{minipage}{0.5\textwidth}
    \centering
    % ===== Plot 3: Fitting error vs r =====
\begin{tikzpicture}
\begin{axis}[
    xlabel={$r$},
    ylabel={Fitting error $\sqrt{\langle (n-f)^2/n^2 \rangle}$},
    grid=major,
]

\addplot[square_color, mark=*, thick]
    table[x=r, y=square, col sep=comma]{fit_bonds/fit_error_vs_r.csv};
%\addlegendentry{Square}

\addplot[tri_color, mark=*, thick]
    table[x=r, y=triangular, col sep=comma]{fit_bonds/fit_error_vs_r.csv};
%\addlegendentry{Triangular}

\addplot[honey_color, mark=*, thick]
    table[x=r, y=honeycomb, col sep=comma]{fit_bonds/fit_error_vs_r.csv};
%\addlegendentry{Honeycomb}

\addplot[corner_color, mark=*, thick]
    table[x=r, y=square_corners, col sep=comma]{fit_bonds/fit_error_vs_r.csv};
%\addlegendentry{Square corners}

\end{axis}
\end{tikzpicture}
    \hfill 
\end{minipage}
    \caption{(a) Numerical counting of $n_\text{legs}$  for various values of $\theta$, fitted to the expected $\theta$-dependence in Eq. (\ref{eq:fit_function}), for a system with radius $r=60a$, where $a$ is the lattice spacing, for three lattice types: square, triangular and honeycomb lattice. For the square lattice, the corners are also counted, in a separate plot.  
    (b) The extracted coefficient $\beta$  of Eq. (\ref{eq:fit_function})  as a function of $r$, is found to remain relatively independent of $r$ (up to a relative error of $\sim10^{-2}$ ), that is, independent of $|\partial A|$, both for the number of virtual legs and for the number of lattice corners crossed by the bipartition boundary. 
    (c) Fitting quality of $n_\text{legs},n_\text{corners}$  to the expected $\theta$-dependence as in Eq. (\ref{eq:fit_function}). 
    In all graphs, the results were obtained by averaging over different orientations of $A$ with resolution $2\pi\cdot10^{-2}$  and different apex positions with resolution $10^{-1}a$  in both dimensions.}
    \label{fig:counting_numerics}
\end{figure}

Finally, since the analytical framework developed in the preceding sections provides an upper bound rather than a closed-form expression for the entanglement corner contribution, we perform a numerical investigation to assess the tightness of this bound. We consider a generalized toric code model with string tension:
\begin{equation}
	\label{eq:orus_model}
    \k{\psi(\gamma)}\propto\prod_l (1+\gamma \sigma^z_l)\k{\psi_0},
\end{equation}
where $l$  denotes the lattice links (edges) and $\k{\psi_0}$ is the toric code ground state with open boundary conditions, $\k{\psi_0}=\prod_p\frac{1+\otimes_{l\in p}\sigma^x_l}{\sqrt{2}}\k{0}^{\otimes N}$, where $p$ denotes the lattice plaquettes.
This is a physically-motivated model, that serves as the ground state of a local Hamiltonian~\citep{Castelnovo} and exhibits a well-characterized quantum phase transition between a topological (deconfined) phase and a trivial (confined) one at $\gamma\approx0.22$~\citep{orus}. 
Here too we simulate a circular system as in Fig. \ref{fig:bipartition} with radius $r=5a$. We compute the von Neumann entropy of a circular section of angle $\theta$, averaged across the lattice orientation $\varphi$ and apex position, and fit the numerical data to Eq. (\ref{eq:fit_function}).
The results are displayed in Fig. \ref{fig:orus_numerics}. 
From Fig. \ref{fig:orus_numerics}a, one may note that the obtained bound appears to be fairly tight in the simulated model, with a slight mismatch for small $\theta$, as expected for a finite lattice size. The topological phase transition, as observed in Ref.~\citep{orus}, is expressed in the fitting parameters $\alpha,\beta$ of Eq. (\ref{eq:fit_function}), as can be seen in Fig. \ref{fig:orus_numerics}b. Note that the bound remains tight on both sides of the phase transition.

\begin{figure}
\raggedright
(a)
  \begin{minipage}{0.5\textwidth}
    \centering
    \input{fit_bonds/orus_entropy_vs_theta}
%    \hfill 
\end{minipage}
(b)
  \begin{minipage}{0.5\textwidth}
    \centering
    \definecolor{RoyalBlue}{HTML}{0000FF}

\begin{tikzpicture}
  \begin{axis}[
    xlabel={$\gamma$},
    ylabel={$\beta$},
    grid=both,
    grid style={line width=0.2pt, draw=gray!30},
    major grid style={line width=0.4pt, draw=gray!45},
    height=4cm,
    width=0.9\linewidth,
    yticklabel style={
        /pgf/number format/fixed,
        /pgf/number format/precision=2
    },
  ]
    \addplot[RoyalBlue, very thick, mark=*, mark size=1.9pt] coordinates {(0.000000,-0.123536) (0.125000,-0.124904) (0.250000,-0.130822) (0.375000,-0.016407) (0.500000,-0.000690) (0.625000,-0.000028) (0.750000,-0.000001) (0.875000,-0.000000) (1.000000,0.000000)};
       \draw[dashed] (axis cs:0.25, -0.14) -- (axis cs:0.25, 0.01);
  \end{axis}
\end{tikzpicture}
%    \hfill 
\end{minipage}
 \begin{minipage}{0.5\textwidth}
    \centering
    % Requires in preamble:
% \usepackage{tikz,pgfplots,xcolor}
% \pgfplotsset{compat=1.18}
\definecolor{RoyalBlue}{HTML}{0000FF}

\begin{tikzpicture}
  \begin{axis}[
    xlabel={$\gamma$},
    ylabel={$\alpha$},
    grid=both,
    grid style={line width=0.2pt, draw=gray!30},
    major grid style={line width=0.4pt, draw=gray!45},
    height=4cm,
    width=0.9\linewidth,
    yticklabel style={
        /pgf/number format/fixed,
        /pgf/number format/precision=2
    },
  ]
    \addplot[RoyalBlue, very thick, mark=*, mark size=1.9pt] coordinates {(0.000000,5.239285) (0.125000,4.936453) (0.250000,3.421617) (0.375000,0.385615) (0.500000,0.021041) (0.625000,0.001022) (0.750000,0.000024) (0.875000,0.000000) (1.000000,0.000000)};
    \draw[dashed] (axis cs:0.25, -0.3) -- (axis cs:0.25, 5.5);
    \end{axis}
\end{tikzpicture}
%    \hfill 
\end{minipage}
    \caption{Numerical fit of the entanglement corner dependence in the toric code ground state with string tension, Eq. (\ref{eq:orus_model}). (a) The von Neumann entropy as a function of $\theta$ for several string tensions $\gamma$. Up to finite-size effects, apparent for small values of $\theta$, the corner-dependence behaves fairly similar to the bound in Eq. (\ref{eq:fit_function}). (b) The coefficients in Eq. (\ref{eq:fit_function}) display a an incipient singularity (rounded off by finite size effects) at the quantum transition,which occurs at $\gamma\approx 0.25$, as marked by dashed lines in the plots~\citep{orus}.}
    \label{fig:orus_numerics}
\end{figure}

\section{Conclusions and future outlook \label{sec:conclusion}}
We studied the subleading corner-dependent term in the entanglement of two-dimensional systems, and specifically, its representability by iPEPSs. This corner term is defined for continuous field systems, while iPEPSs represent discretized systems, whether they originally reside on a discretized lattice or serve as an approximation of a continuous system.

We show, based on the bond dimension of the iPEPS and its relation to entanglement, that in the continuum limit a corner-dependent and area-independent term emerges, similar to the one predicted in continuous gapped systems. A similar argument shows that such a behavior may also rise from the lattice-corner-dependence of the entanglement in gauge-invariant systems. Our result ties the entanglement behavior in continuous systems with the geometrical behavior of its discretized version in the general case. It could be useful to compare this viewpoint to the one presented in Ref.~\citep{PokManTamCornerCharge2024} for the charge fluctuations in free fermion systems, which is related there to the integrated quantum metric. 
Trading accuracy for broad applicability, we establish a universal geometric bound on the entanglement that, despite being an inequality, is shown numerically to be remarkably tight throughout a quantum phase transition.

The technical aspects of our result lead to an important practical insight: the behavior observed in the continuum could not have emerged from a single fixed lattice orientation $\varphi$. It required averaging over all possible orientations to restore the system's isotropy. Similarly, this term would not appear if the discretized lattice was not translational invariant, e.g., if it were centered with respect to the system's center. This emphasizes the necessity of such symmetrization procedures when approximating continuous systems by discretization.

In hindsight, the connection between entanglement in continuous systems and the geometry of their discretized counterparts is perhaps not surprising: The definition of entanglement in such systems requires the introduction of an energy cutoff, that is, a discretization of the system~\citep{casini2023lecturesentanglementquantumfield}. It would be interesting to explore whether other entanglement properties, across different spatial dimensions, can similarly be linked to geometric features. Extending the argument to critical systems with long correlation lengths also presents a compelling direction for future investigation.

 \section{Acknowledgements}
We thank E. Bettelheim, P.M. Tam, and E. Zohar for fruitful discussions. Our work has been supported by the Israel Science Foundation (ISF) and the Directorate for Defense Research and Development (DDR\&D) through Grant No. 3427/21, the ISF Grant No. 1113/23, and the US-Israel Binational Science Foundation (BSF) through Grant No. 2020072. N.F. is also supported by the Azrieli Foundation Fellows program and the Milner Foundation.

\appendix*
\section*{Appendix: Corner dependence of the bond dimension when $\theta\rightarrow\pi$}
\label{sec:appendix_large_theta}
\phantomsection % Create a valid link target
% \makeatletter
% \addcontentsline{toc}{section}{Appendix} % Optional: adds it to the TOC
% \def\@currentlabelname{Appendix}% Manually set the name
% \makeatother

% \section{}

\begin{figure}[t]
    \centering
    \includegraphics[width=\linewidth]{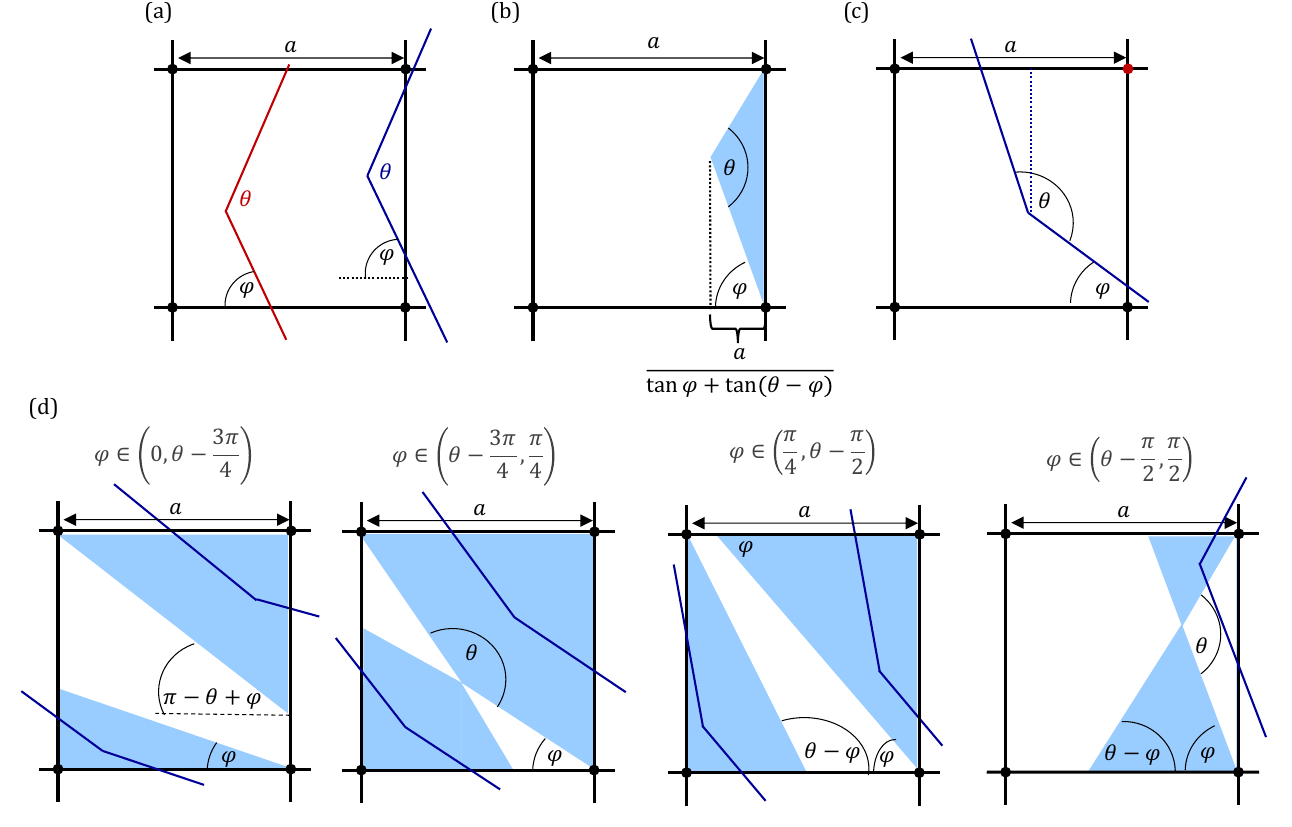}
    \caption{(a) When $\theta\rightarrow\pi$, its effect on $n_\text{legs}$  stems from the bond legs closest to the vertex. We examine a corner with an opening angle $\theta$  and an orientation angle $\varphi$. Different positions of the vertex result in different contributions to $n_\text{legs}$. The red edges crosses the top and bottom bonds, while the blue edges do not pick up any lattice site in the unit cell, i.e., no lattice site is included in $A$. Thus, the edges do not cross any bond. (b) For given $\theta,\varphi$, the blue region denotes the vertex positions which lead to skipping of a bond.  Its area may be geometrically shown to be $a^2/2(\tan\left(\theta-\varphi)+\tan\varphi\right)$. (c) When $\varphi < \theta-\pi/2$, it is geometrically apparent that at least on of the lattice sites is included in $A$ (see the site colored in red). Thus, $p_\text{skip}(\theta, \varphi)=0$ for these orientations. (d) For given $\theta,\varphi$, the unit cell regions which contribute to $n_\text{corners}$  are colored in light blue. Example edges are added in dark blue for each of the cases.    }
    \label{fig:appendix}
\end{figure}

In this Appendix, we obtain explicitly the dependence of $n_\text{legs}, n_\text{corners}$ on $\theta$  for $\theta\rightarrow\pi$, and show that the dependence matches Eq. (\ref{eq:theta_dependence}) as in continuum field theory.

Recall that we average $n_\text{legs},n_\text{corners}$ over the orientation of subsystem $A$ denoted by $\varphi$, and over the vertex position in the unit cell. We start with the dependence of $n_\text{legs}$ on $\theta$. When $\theta\rightarrow\pi$, its effect on $n_\text{legs}$  stems from lattice sites closest to the vertex, which may be skipped or picked up by the boundary, depending on $\theta$, see Fig. \ref{fig:appendix}a. For a given corner $\theta$  and orientation $\varphi$ and varying vertex position, the probability of skipping the nearest lattice sites (see the geometric depiction in Fig. \ref{fig:appendix}b):
\begin{equation}\label{eq:pskip_varphi}
    p_\text{skip}(\theta,\varphi) =  \frac{1}{2(\tan(\theta-\varphi)+\tan\varphi)}.
\end{equation}
We now integrate over the orientation $\varphi$. Note that the expression above is positive for $\varphi - m\pi/2\in(\theta-\pi/2,\pi/2)$ where $m$  is an integer. For orientations outside of this range, $p_\text{skip}$  may be geometrically shown to be 0, as illustrated  in Fig. \ref{fig:appendix}c.  The final expression is thus obtained, 
\begin{equation}
    p_\text{skip}(\theta)=\int_{\theta-\pi/2}^{\pi/2}p_\text{skip}(\theta,\varphi)d\varphi= \frac{1}{4}\left(1-(\pi-\theta)\cot\theta\right),
    \label{eq:pskip}
\end{equation}
with a corner dependence identical to the continuous system behavior in Eq. (\ref{eq:theta_dependence}). Note that $p_\mathrm{skip}(\theta)$ contributes to the term subtracted from the entropy, and therefore it acquires a minus sign when substituted in Eq. \ref{eq:theta_dependence}. The above analysis applies for all $\theta$s large enough such that they will surely pick up the bonds in the neighboring plaquette, which in a square lattice requires $\theta<2\tan^{-1}(1/2)\approx0.3\pi$. Therefore, Eq. (\ref{eq:pskip}) applies to the range $0.3\pi<\theta < \pi$, and not only for $\theta\rightarrow\pi$.

We now turn to on $n_\text{corners}$. In Fig. \ref{fig:appendix}d, we depict the vertex positions for which a boundary corner $\theta$  of orientation $\varphi$  picks up a corner of the lattice. Here we consider different ranges of orientations $\varphi$. Let us start from the probability to pick up a corner when $\varphi\in(\theta-\pi/2,\pi/2)$, which is again obtained geometrically:
\begin{equation}
    p_\text{corner}(\theta,\varphi)=\frac{1}{2}\cot\varphi + \frac{1}{2}\cot(\theta-\varphi)-2p_\text{skip}(\theta,\varphi).
\end{equation}
Integrating over the orientations, one obtains
\begin{equation}
    p_\text{corner}(\theta)=\int_{\theta-\pi/2}^{\pi/2}p_\text{corner}=(\theta,\varphi)d\varphi= \log\cos(\pi-\theta) - 2p_\text{skip}(\theta).
\end{equation}
Since $\theta\rightarrow\pi$, we find
\begin{equation}
    p_\text{corner}(\theta)=-2p_\text{skip}(\theta)  + O\left((\pi-\theta)^2\right),
    \end{equation}
    recovering the behavior of Eq. (\ref{eq:theta_dependence}), similarly to $n_\text{legs}$. Equivalent calculations may be performed for the remaining values of $\varphi$ to obtain a similar dependence on $\pi-\theta$. 

\bibliographystyle{apsrev4-2}
\bibliography{ipeps_corner}

\end{document}